\documentclass{article}
\usepackage{amssymb}
\usepackage{amsmath}

\input{tcilatex}
\begin{document}

\title{\ \textbf{Quantum Gravity and a Time Operator in Relativistic Quantum
Mechanics}}
\author{\textit{M. Bauer} \\
Instituto de F\'{\i}sica\\
Universidad Nacional Aut\'{o}noma de M\'{e}xico\\
bauer@fisica.unam.mx}
\maketitle

\begin{abstract}
The problem of time in the quantization of gravity arises from the fact that
time in Schr\"{o}dinger's equation is a parameter. This sets time apart from
the spatial coordinates, represented by operators in quantum mechanics (QM).
Thus "time" in QM and "time" in General Relativity (GR) are seen as mutually
incompatible notions. The introduction of a dynamical time operator in
relativistic quantum mechanics (RQM), that in the Heisenberg representation
is also a function of the parameter t (identified as the laboratory time),
prompts to examine whether it can help to solve the disfunction referred to
above. In particular, its application to the conditional interpretation of
the canonical quantization approach toquantum gravity is developed.
\end{abstract}

\section{Introduction}

Quantization of general relativity (GR) is still an unsolved problem in
physics. One of the difficulties, referred as the \textit{problem of time},
arises from the fact that time in quantum mechanics (QM) is a parameter\cite%
{Dirac,Pauli}. The extensive experimental confirmation of the Schr\"{o}%
dinger equation identifies this parameter as the laboratory time, thus part
of the the space--time frame of reference associated to an observer. This
sets time apart from the system's physical properties (e.g., energy,
momentum, position), which are represented by operators, whereas time is not%
\footnote{%
In this respect it is important to avoid, following Hilgevoord\cite%
{Hilgevoord1}, the confusion between the space coordinates $(x,y,z)$ of a
system of reference and the quantum mechanical position operator $\mathbf{%
\hat{r}=(}\hat{x},\hat{y},\hat{z})$\ whose expectation value gives the time
evolution of the position of a system described by a certain state vector $%
\mid \Psi (t)>$. In the same way, a distinction should be made between the
time coordinate in $(x,y,z,t)$ and a "time operator $\hat{T}$" acting on the
system state vector. The space coordinates $(x,y,z)$ together with the
parameter $t$ constitute the background spacetime framework subject to
Lorentz transformations in special relativity, whereas the operators $%
\mathbf{\hat{r}=(}\hat{x},\hat{y},\hat{z})$ and, if it exists, $\hat{T}$ ,
are to be considered as, say, "path" dynamical space and time operators.}.\
On the other hand, in general relativity (GR) where matter determines the
structure of spacetime, time and space acquire a dynamical nature. Thus
"time" in QM and the "time" in GR are seen as mutually incompatible notions%
\cite{Anderson,Kushar,Kiefer,Macias,Isham,Ashtekar}.

Quantization of general relativity has been approached in different ways.
The canonical quantization approach, to be considered here, is based on a
3+1 decomposition of spacetime, namely a foliation of three-dimension
spacelike hypersurfaces and a one-dimension timelike vector that may
characterize the foliation. The Dirac prescription to transform dynamical
variables into operators, as used to formulate standard quantum mechanics
from the Hamiltonian-Jacobi formulation of classical mechanics, is then
applied\cite{DeWitt, Anderson,Kushar,Kiefer,Isham}. It results, however, in
the Wheeler-de Witt equation (WdW), where time is absent. The problem of
time is that there is no time. The WdW equation predicts a static universe,
contrary to obvious everyday experience.

To resolve this contradiction, Page and Wooters (PW)\cite%
{Page,Dolby,Giovaneti} advanced that a static system may describe an
evolving "universe" from the point of view of an internal observer, by
introducing conditional probabilities between two of the system observables,
the continuum spectrum of one of them serving as the "internal time"
parameter for the other. However, as in ordinary Schr\"{o}dinger quantum
mechanics the probability amplitudes of all dynamical variables are referred
to a single time\footnote{%
Indeed, as pointed out in Ref.(13), \textquotedblleft a consistent
definition of a probability density can include only points on a space-like
surface, i.e., with no possible causal connection. In the non-relativistic
limit $(c=\infty )$\ all such surfaces are reduced to $\tau =const$\ planes,
and the normalization applies only to the domain of space dimensions. Thus
under no circumstances is the time variable on a complete equal footing as
the space variables.\textquotedblright\ }, it was soon pointed out that the
chosen observable should condition not one but all other dynamical variables
and, to quote\cite{Unruh}: "Evidence against the possibility of using a
dynamical variable to play the role of "time" in the conditional probability
interpretation is provided by the fact (proven here) that in ordinary Schr%
\"{o}dinger quantum mechanics for a system with a Hamiltonian bounded from
below, no dynamical variable can correlate monotonically with the Schr\"{o}%
dinger parameter $t$, and thus the role of $t$\ in the interpretation of Schr%
\"{o}dinger quantum mechanics cannot be replaced by that of a dynamical
variable". This is the well known objection to the existence of a time
operator in quantum mechanics, raised by Pauli\cite{Pauli}. \ To be noted
finally is that the non existence of a time operator seems to have been
taken for granted in all up to date developments in quantum gravity\cite%
{Anderson,Kushar,Kiefer,Macias,Isham,Ashtekar}.

In the present paper, however, the introduction of a self-adjoint dynamical
time operator in relativistic quantum mechanics (RQM)\cite{Bauer}, that in
the Heisenberg picture is also a function of the parameter $t$ , prompts to
examine whether it can help to solve the disfunction referred to above, as
well as support the conditional probability interpretation of canonical
quantum gravity.

In Section 2, the proposed self-adjoint dynamical time operator in Dirac's
relativistic quantum mechanics ($\hat{T}=\mathbf{\alpha .\hat{r}}/c$ +$\beta
\tau _{0})$ is presented in addition to the usual dynamical dynamical
observables. It does provide a time energy uncertainty relation related to
the position momentum one, as surmised by Bohr, and circumvents Pauli's
objection by giving rise to energy changes through momentum displacements.
In Section 3 the properties that the conditionning PW operator has to
satisfy are reviewed and shown how they can be fulfilled by the time
operator. Conclusions and further possible applications are presented
finally.

\section{The dynamical time operator\protect\cite{Bauer}}

In analogy to the Dirac free particle Hamiltonian $\hat{H}_{D}=c\mathbf{%
\alpha .\hat{p}}+\beta m_{0}c^{2\text{ }}$where $\alpha _{i}(i=1,2,3)$ and $%
\beta $ are the $4\times 4$ Dirac matrices, a dynamical "time operator" $%
\hat{T}=\mathbf{\alpha .\hat{r}/}c\mathbf{+}\beta \tau _{0\text{ }}$has been
introduced. It is shown that:

A) its eigenvectors are 
\begin{equation}
\left\vert \tau \right\rangle =u_{r}\left\vert \mathbf{r}\right\rangle
\end{equation}%
where $\left\vert \mathbf{r}\right\rangle $ is the eigevector of the
position operator and $u_{r}$\ is a four component spinor independent of the
linear momentum $\mathbf{p}$. The corresponding eigenvalues are:%
\begin{equation}
\tau =\pm \lbrack (r/c)^{2}+\tau _{0\text{ }}^{2}]^{1/2}
\end{equation}%
There is a continous positive and a continous negative "time branch". As $[%
\mathbf{\sigma .\hat{r}}/2r,\hat{T}]=0$, where $\mathbf{\sigma .\hat{r}}/2r$
represents the spin component in the direction of $\mathbf{r}$, there are
four independent "time spinors", corresponding to two spin projections for
each branch. The \ value of the intrinsic property $\tau _{0\text{ }}$is
found to be $h/m_{0}c^{2\text{ }}$, the de Broglie period(Appendix A).\cite%
{Bauer3}

B) in the Heisenberg picture%
\begin{equation}
i\hslash \frac{d\hat{T}}{dt}=[\hat{T},\hat{H}_{D}]=i\hslash \{\mathbf{I}%
+\beta K\}+2\beta \{\tau _{0}\hat{H}_{D}-m_{0}c^{2}\hat{T}\}
\end{equation}%
where $K=\beta (2\mathbf{s.l}/\hslash ^{2}+1)$\ is a constant of motion\cite%
{Thaller}. Integrating one obtains:%
\begin{equation}
\hat{T}(t)=\{1+\beta K\}t+oscillating\ terms
\end{equation}%
\ \ \ Thus the time operator correlates monotonically with the time
parameter $t$ of the Sch\"{o}dinger equation $(\tau \propto t)$.

C) this operator is clearly self-adjoint and therefor can be the generator
of a unitary transformation (Stone's theorem)%
\begin{equation}
U_{T}(\varepsilon )=e^{-i\varepsilon \hat{T}/\hslash }=e^{-i\varepsilon \{%
\mathbf{\alpha .\hat{r}/}c\mathbf{+}\beta \tau _{0}\}/\hslash }
\end{equation}%
where $\varepsilon $\ has the dimensions of energy. For infinitesimal
transformations ($\delta \varepsilon <<1$), the transformed Hamiltonian $%
\tilde{H}_{D}=U\hat{H}_{D}U^{^{\dagger }}$\ is approximated as:%
\begin{equation}
\tilde{H}_{D}=U\hat{H}_{D}U^{^{\dagger }}\simeq (1-i\varepsilon \hat{T}%
/\hslash +...)H_{D}(1+i\varepsilon \hat{T}/\hslash +...)\simeq
H_{D}-i(\varepsilon /\hslash )[\hat{T},\hat{H}_{D}]+...
\end{equation}%
Now:%
\begin{equation}
\lbrack \hat{T},\hat{H}_{D}]=i\hbar \{3\mathbf{I}+4\mathbf{s.l}/\hslash
^{2}\}+2\beta \{\tau _{0}\hat{H}_{D}-m_{0}c^{2}\hat{T}\}.
\end{equation}%
that\ reduces to $[\hat{T},\hat{H}_{D}]=i\hbar 3\mathbf{I}$ in the rest
frame $(\mathbf{r}=0,\mathbf{p}=0,\mathbf{l}=0$ , $\hat{H}_{D}=\beta
m_{0}c^{2}$ and $T=\beta \tau _{0})$. Otherwise there will be a \textit{%
transient} Zitterbewgung behavior about the monotonic evolution\cite%
{Thaller2}

Then, using $\mathbf{\alpha .\alpha =}$ $3\mathbf{I}$, one has:%
\begin{equation}
\tilde{H}_{D}=U\hat{H}_{D}U^{^{\dagger }}=\hat{H}_{D}(\mathbf{p}%
)+\varepsilon \mathbf{\alpha .\alpha =}c\mathbf{\alpha .(\hat{p}+}%
\varepsilon \mathbf{\alpha /}c)+\beta m_{0}c^{2}=\hat{H}_{D}(\mathbf{\hat{p}+%
}\varepsilon \mathbf{\alpha /}c)
\end{equation}%
The unitary transformation induces a shift in momentum (Lorentz boost):%
\begin{equation}
\delta \mathbf{p=\{}(\delta \varepsilon )/c\}\mathbf{\alpha =\{}(\delta
\varepsilon )/c^{2}\}c\mathbf{\alpha }
\end{equation}%
Repeated infinitessimal applications yield a momentum displacement $\mathbf{p%
}$ whose expectation value is%
\begin{equation}
<\ \mathbf{p>}=(\varepsilon /c^{2})\mathbf{v}_{gp}=\gamma m_{0}\mathbf{v}%
_{gp}
\end{equation}%
where $\gamma =\{1-(v_{gp}/c)^{2}\}^{-1/2}$ is the Lorentz factor and $%
\mathbf{v}_{gp}$ the group velocity.

Clearly the shift in momentum represents a displacement in energy within the
positive and the negative energy branches. In the positive branch, as $p$
goes\ from $-\infty $\ to $+\infty $ in any direction, the energy drops from 
$+\infty $\ to a minimun $m_{0}c^{2}$\ at $p=0$ and ther rises again to $%
+\infty $\ .\ No crossing of the $2m_{0}c^{2}$\ energy gap is involved. The
time operator is thus the generator of a unitary transformation that
corresponds to a change in energy. However, by acting on the momentum
continuum space, it circumvents Pauli's objection.

Finally it can be pointed out that, as $\hat{T}$ and $\mathbf{\hat{r}}$
commute, the previous development applies also in the presence of any
position dependent potentials, e.g., the scalar and vector electromagnetic
potentials.

\section{The dynamical time operator and the conditional probability
interpretation of quantum gravity}

Following Ref.5 (Kuchar), the PW\ conditional probability interpretation
asserts that, given $\hat{B}$ and $\hat{C}$ as projection operators
corresponding to observables of the system: 
\begin{equation}
P\left( B\mid C\right) =\frac{\left\langle \Psi \mid \hat{C}\hat{B}\hat{C}%
\mid \Psi \right\rangle }{\left\langle \Psi \mid \hat{C}\mid \Psi
\right\rangle }
\end{equation}%
represents the probability that the observation of $\hat{B}$ is subject to
the observation of $\hat{C}$. It is then stated that the connection with the
time problem is established by finding within the system a projection
operator $\hat{C}(t)$ corresponding to the question "Does an internal clock
show the time $t$?". The operator $\hat{C}(t)$ cannot commute with the
Hamiltonian as otherwise it would be constant. If $[\hat{B},\hat{C}(t)]=0$
the complete Hamiltonian can be written as%
\begin{equation}
\hat{H}=(\hat{h}_{B}\otimes I_{C})\otimes (I_{B}\otimes \hat{h}_{C})
\end{equation}%
in the space $F_{B}\otimes $ $F_{C}$ composed of the corresponding Hilbert
spaces. Then:%
\begin{equation}
\lbrack \hat{H},\hat{C}(t)]=[\hat{h}_{C},\hat{C}(t)]\neq 0
\end{equation}%
Now, if $\left\vert \phi _{0}\right\rangle \ $is\ a\ state\ vector$\ $such\
that$\ \hat{C}(0)=\left\vert \phi _{0}\right\rangle \left\langle \phi
_{0}\right\vert $ corresponds to $t=0$, it follows that%
\begin{equation}
\hat{C}(t)=e^{i\hat{H}t/\hslash }\hat{C}(0)e^{-i\hat{H}t/\hslash }=e^{i\hat{h%
}_{C}t/\hslash }\hat{C}(0)e^{-i\hat{h}_{C}t/\hslash }
\end{equation}%
Then:%
\begin{eqnarray}
P(B|C(t)) &=&\frac{\left\langle \Psi \mid e^{i\hat{h}_{C}t/\hslash
}\left\vert \phi _{0}\right\rangle \left\langle \phi _{0}\right\vert e^{-i%
\hat{h}_{C}t/\hslash }\hat{B}e^{i\hat{h}_{C}t/\hslash }\left\vert \phi
_{0}\right\rangle \left\langle \phi _{0}\right\vert e^{-i\hat{h}%
_{C}t/\hslash }\mid \Psi \right\rangle }{\left\langle \Psi \left\vert \phi
_{0}\right\rangle \left\langle \phi _{0}\right\vert \Psi \right\rangle } 
\notag \\
&=&\frac{\left\langle \psi \mid \hat{B}(t)\mid \psi \right\rangle }{%
\left\langle \psi \mid \psi \right\rangle }
\end{eqnarray}%
where $\left\vert \psi \right\rangle =\left\langle \phi _{0}\mid \Psi
\right\rangle $ and:%
\begin{equation}
\hat{B}(t)=e^{-i\hat{h}_{B}t/\hslash }\hat{B}e^{i\hat{h}_{B}t/\hslash }=e^{-i%
\hat{H}t/\hslash }\hat{B}e^{i\hat{H}t/\hslash }
\end{equation}%
It follows then that $\hat{B}$ also satisfies:%
\begin{equation}
i\hslash \frac{d\hat{B}}{dt}=[\hat{B},\hat{h}_{B}]=[\hat{B},\hat{H}]
\end{equation}%
in spite of the fact that $\left\vert \Psi \right\rangle $is a stationary
state of the total system.

Note that this development is already assuming a time dependent Schr\"{o}%
dinger equation\ (TDSE), without explaining the presence in it of the
laboratory time $t$.This will be addressed below.

The connection of the PW conditional interpretation with the dynamical time
operator defined above is as follows. Besides being a timelike operator as
it is given in terms of the worldline $\mathbf{r}(t)$, $\hat{T}$ does
satisfy the following conditions:

i) it does not commute with the Hamiltonian;

ii) its spectrum is a single valued continuum in either positive or negative
branch, directly proportional to the time parameter $t$ ;

iii) the eigenvector basis $\{\left\vert \tau \right\rangle =u_{r}\left\vert 
\mathbf{r}\right\rangle \}$\ where $\tau \propto t$\ can be used to
construct the normalized wave packet $\left\vert \phi _{0}\right\rangle
=\int d\tau c_{\tau }\left\vert \tau \right\rangle $\ such that%
\begin{equation}
\left\langle \phi _{0}\right\vert \hat{T}\left\vert \phi _{0}\right\rangle
=\int \int d\tau d\tau ^{^{,}}c_{\tau ^{\prime }}c_{\tau }\left\langle \tau
^{\prime }\right\vert \hat{T}\left\vert \tau \right\rangle =\int d\tau
\left\vert c_{\tau }\right\vert ^{2}\tau \propto \int dt\left\vert
c_{t}\right\vert ^{2}t=0
\end{equation}

iv) as $[\hat{T},\mathbf{\hat{r}}]=0$ , one can consider:%
\begin{equation}
\hat{B}=\left\vert \mathbf{r}\right\rangle \left\langle \mathbf{r}\right\vert
\end{equation}

Then, with $\Psi (\mathbf{r},\tau ):=\left\langle \Psi \left\vert \tau
\right\rangle \left\langle \tau \right\vert \mathbf{r}><\mathbf{r}\left\vert
\tau \right\rangle \left\langle \tau \right\vert \Psi \right\rangle ,:$%
\begin{equation}
P(B\mid C)=\frac{\left\langle \Psi \left\vert \tau \right\rangle
\left\langle \tau \right\vert \mathbf{r}><\mathbf{r}\left\vert \tau
\right\rangle \left\langle \tau \right\vert \Psi \right\rangle }{%
\left\langle \Psi \mid \tau ><\tau \mid \Psi \right\rangle }=\frac{%
\left\vert \Psi (\mathbf{r},\tau )\right\vert ^{2}}{\int d\mathbf{r}%
\left\vert \Psi (\mathbf{r},\tau )\right\vert ^{2}}
\end{equation}%
would be the probability density for finding the system at value $\mathbf{r(}%
t\mathbf{)}$ at an instant $\tau (t)\propto t$ .

\bigskip

The eigenvectors $\mid \tau ,\mathbf{r}>$ (common eigenvectors of $\mathbf{r}
$ and $T$ as $[T,\mathbf{r}]=0$) constitute a basis. In the Sch\"{o}dinger
picture, they give an "intrinsic time"-space spinor representation $\Psi
(\tau ,\mathbf{r};t)=<\tau ,\mathbf{r}\mid \Psi (t)>$ of the time dependent
Schr\"{o}dinger state vector. This is entirely analogue to the
energy-momentum spinor representation $\Phi (E,\mathbf{p};t)=<E,\mathbf{p}%
\mid \Psi (t)>$ where \ $\mid E,\mathbf{p}>$ are the common eigenvectors of
the relativistic free particle Dirac Hamiltonian $H_{D}=c\mathbf{\alpha .p}%
+\beta m_{0}c^{2}$ and the momentum operator $\mathbf{p}$ \textbf{. }The
time dependence of $T$ is exhibited in the Heisenberg picture and seen to
correlate monotonically with the parameter $t$ for wave packets of purely
positive (or purely negative)$\ \tau $ eigenstates\cite{Bauer}

Consequently $\left\vert \Psi (\tau ,\mathbf{r};t)\right\vert ^{2}$ is
interpreted as the probability of finding at time $t$ the system at position 
$\mathbf{r}$ and intrinsic time $\tau =\pm \sqrt{(r/c)^{2}+\tau _{0}^{2}}.$
Normalization of $\Psi (\tau ,\mathbf{r};t)$ includes sum over spin but no
integration over an extra dimension beyond $\mathbf{r}$\textbf{\cite{Greiner}%
}. Then one has:%
\begin{equation}
P(\tau ,\mathbf{r;}t\mathbf{)=}\frac{\left\vert \Psi (\tau ,\mathbf{r}%
;t)\right\vert ^{2}}{\dsum\limits_{\alpha }\dint d\mathbf{r}\left\vert \Psi
(\tau ,\mathbf{r};t)\right\vert ^{2}}
\end{equation}

\bigskip

\section{ Conclusion}

The introduction of a self-adjoint time operator in RQM, in addition to the
usual dynamical variables, allows to consider its possible role in the
"problem of time" in quantum gravity (QG). As defined, this time operator
has a one to one correspondence with the timelike worldline $r(t)$\ .\ Then
to each point of its spectrum one can associate a spacelike surface that
intersects the worldline at the corresponding point, thus providing a
foliation of spacetime by spacelike surfaces over which one can define
probability amplitudes\textit{. }Furthemore it provides support to the
conditional probability interpretation of PW of the canonical quantization
of QG, by circumventing Pauli's objection to the existance of such an
operator, as well as providing a monotonical correlation with the time
parameter in the Schr\"{o}dinger equation. Consequently one can say that
this operator yields an observable dynamical variable that \textquotedblleft
sets the conditions\textquotedblright\ for the other variables and defines a
satisfactory notion of time.

It is interesting to note that the presence of the parameter $t$ in the time
dependent Schr\"{o}dinger equation can be attributed to the \textit{%
monitoring} that a classical environment, interacting with the microscopic
system, exerts on the system. Indeed, starting from a time independent Schr%
\"{o}dinger equation with a complete Hamiltonian (system, environment and
interaction), the system is shown to satisfy a time dependent Schr\"{o}%
dinger equation when it is disentangled from its classically described
(dependent on the laboratory time $t$) environment. To quote: "The time
dependence (- and perhaps also the space dependence, conforming together the
Minkowskian spacetime laboratory reference frame) - is thus seen as an
emergent property, both in QM and in QG" \textit{\cite{Briggs,Butterfield}. }%
Furthermore an intermediate subdivision can be introduced\cite{Briggs2},
that in our case allows the presence of two times, a system "internal time"
constructed by the Page Wooters mechanism, and the "laboratory time" arising
from the interaction with a massive classical environment. An experimental
illustration of, to quote: "A static, entangled system between a clock
system and the rest of the universe is perceived as evolving by internal
observers that test the correlations between the two subsystems"\textit{\ }%
has allready been achieved\cite{Moreva}

The presentation in this paper is at a basic level, as the stress is on the
fact that a dynamical time operator in RQM can be defined, contrary to the
general view. It remains to be formulated in the usual 3+1 foliation of the
spacetime with Riemann spacelike sufaces, and, as it introduces "time
spinors", its possible relevance to the more advanced formulation of loop
quantum gravity.

\bigskip

--------------------------------------------------------------------------

\textbf{Appendix A}

For infinitesimal transformations ($\delta \varepsilon <<1$), one can
factorize the unitary operator$U_{T}(\varepsilon )$ generated by the time
operator as follows :%
\begin{equation}
U_{T}(\varepsilon )\simeq e^{-i(\delta \varepsilon )\{\mathbf{\alpha .\hat{r}%
}/c\}/\hbar }e^{-i(\delta \varepsilon )\beta \tau _{0}/\hbar }=e^{-i(\delta
\varepsilon )\beta \tau _{0}/\hbar }e^{-i(\delta \varepsilon )\{\mathbf{%
\alpha .\hat{r}}/c\}/\hbar }  \tag{A.1}
\end{equation}%
as $\ [i(\delta \varepsilon )(\mathbf{\alpha .r}/c\hbar ),i(\delta
\varepsilon )\beta \tau _{0}/\hbar ]\approx (\delta \varepsilon )^{2}\approx
0$ (Glauber theorem). Then the transformed Hamiltonian can be approximated
as: 
\begin{equation}
\tilde{H}_{D}=U\hat{H}_{D}U^{^{\dagger }}\simeq e^{i(\delta \varepsilon
)\beta m_{0}c^{2}/\hslash }e^{i(\delta \varepsilon )\mathbf{\alpha .\hat{r}}%
/c\hslash }\hat{H}_{D}e^{-i(\delta \varepsilon )\mathbf{\alpha .\hat{r}}%
/c\hslash }e^{-i(\delta \varepsilon )\beta m_{0}c^{2}/\hslash }  \tag{A.2}
\end{equation}

Consider first:

\begin{eqnarray}
&&e^{i(\delta \varepsilon )\mathbf{\alpha .\hat{r}}/c\hslash }\hat{H}%
_{D}e^{-i(\delta \varepsilon )\mathbf{\alpha .\hat{r}}/c\hslash }  \notag \\
&\simeq &\{I+i(\delta \varepsilon )\mathbf{\alpha .\hat{r}}/c\hslash +...\}%
\hat{H}_{D}\{I-i(\delta \varepsilon )\mathbf{\alpha .\hat{r}}/c\hslash +...\}
\notag \\
&\simeq &\hat{H}_{D}+i\{(\delta \varepsilon )/c\hslash \}[\mathbf{\alpha .%
\hat{r},}\hat{H}_{D}]+....  \TCItag{A.3}
\end{eqnarray}%
Then using\cite{Thaller}:%
\begin{equation}
\lbrack \hat{H}_{D},\mathbf{\alpha .\hat{r}]}\mathbf{=}-3i\mathbf{\hslash }c%
\mathbf{I+}2\hat{H}_{D}\{\mathbf{\alpha }-c\mathbf{p}/H_{D}\}.\mathbf{\hat{r}%
}==-i\mathbf{\hslash }c\mathbf{\alpha .\alpha +}2\hat{H}_{D}\mathbf{\alpha .%
\hat{r}}-2c\mathbf{p}.\mathbf{\hat{r}}  \tag{A.4}
\end{equation}%
and $\mathbf{\alpha .\alpha =}3\mathbf{I}$, one obtains:%
\begin{equation}
e^{i(\delta \varepsilon )\mathbf{\alpha .\hat{r}}/c\hslash }\hat{H}%
_{D}e^{-i(\delta \varepsilon )\mathbf{\alpha .\hat{r}}/c\hslash }\simeq \hat{%
H}_{D}(\mathbf{p+}\delta \varepsilon \mathbf{\alpha /}c\mathbf{)+}i2\{\delta
\varepsilon /c\hslash \}\{\hat{H}_{D}\mathbf{\alpha .\hat{r}}-c\mathbf{p}.%
\mathbf{\hat{r}\}}  \tag{A.5}
\end{equation}%
Thus, the unitary transformation induces a shift in momentum:%
\begin{equation}
\delta \mathbf{p=\{}(\delta \varepsilon )/c\}\mathbf{\alpha =\{}(\delta
\varepsilon )/c^{2}\}c\mathbf{\alpha }  \tag{A.6}
\end{equation}%
as well as a Zitterbewegung behavior in the corresponding propagator $%
U(t)=e^{-i\tilde{H}_{D}t/\hslash }$.

For repeated infinitessimal applications one obtains a momentum displacement 
$\mathbf{p}$ whose expectation value is%
\begin{equation}
<\ \mathbf{p>}=(\varepsilon /c^{2})\mathbf{v}_{gp}=\gamma m_{0}\mathbf{v}%
_{gp}  \tag{A.7}
\end{equation}%
where $\gamma =\{1-(v_{gp}/c)^{2}\}^{-1/2}$ is the Lorentz factor and $%
\mathbf{v}_{gp}$ the group velocity. It also induces a phase shift. Indeed:%
\begin{equation}
\left\langle \Psi \right\vert \tilde{H}_{D}\left\vert \Psi \right\rangle
=\left\langle \Phi \right\vert \hat{H}_{D}(\mathbf{p+\alpha }\text{ }\delta
\varepsilon /c)\left\vert \Phi \right\rangle =\left\langle \Phi \right\vert 
\hat{H}_{D}(\mathbf{p+}\gamma m_{0}\mathbf{\mathbf{v}_{gp}})\left\vert \Phi
\right\rangle  \tag{A.8}
\end{equation}%
where%
\begin{equation}
\left\vert \Phi \right\rangle =e^{-i(\delta \varepsilon )\beta \tau
_{0}/\hslash }\left\vert \Psi \right\rangle  \tag{A.9}
\end{equation}%
The phase shift is $\ \delta \varphi =-(\delta \varepsilon )\beta \tau
_{0}/\hslash \ $. For a finite transformation, its expectation value is%
\begin{equation}
\left\langle \Delta \varphi \right\rangle =-\{(\Delta \varepsilon )\tau
_{0}/\hslash \}\ \left\langle \beta \right\rangle =\mp m_{0}c^{2}\tau
_{0}/\hslash )  \tag{A.10}
\end{equation}%
as $<\beta >=m_{0}c^{2}/<H_{D}>=\pm m_{0}c^{2}/\varepsilon =\pm 1/\gamma $ ,
for a positive (negative) energy wave packet that contains both positive and
negative energy free particle solutions\cite{Thaller}.Thus the sign of $%
<\beta >$ distinguishes the positive or negative energy branch where the
momentum displacement takes place. If furthermore one requires the
corresponding phase shift to be equal to $2\pi $ , one has to set:%
\begin{equation}
\tau _{0}=2\pi \hbar /<\beta >\varepsilon =h/m_{0}c^{2}  \tag{A.11}
\end{equation}%
This is the de Broglie period. One has then:%
\begin{equation}
h/p=h/\gamma m_{0}v_{gp}=hc^{2}/\gamma m_{0}c^{2}v_{gp}=(h/\varepsilon
)(c^{2}/v_{gp})=(1/\nu )v_{ph}  \tag{A.12}
\end{equation}%
which is precisely the de Broglie wave length, that is, the product of the
phase velocity by the period derived from the Planck relation $E=h\upsilon $%
\ and the Einstein relation $E=m_{0}c^{2}$\ , as originally assumed by de
Broglie.\cite{Broglie}

As the state vector $\left\vert \Phi \right\rangle $ differs from the state
vector $\left\vert \Psi \right\rangle $ by a global phase, it follows that:%
\begin{equation}
\left\langle \Phi \right\vert \hat{H}_{D}(\mathbf{p+}\gamma m_{0}\mathbf{%
\mathbf{v}_{gp}})\left\vert \Phi \right\rangle =\left\langle \Psi
\right\vert \hat{H}_{D}(\mathbf{p+}\gamma m_{0}\mathbf{\mathbf{v}_{gp}}%
)\left\vert \Psi \right\rangle  \tag{A.13}
\end{equation}

Finally, it is also interesting to note the following. In the same way as
above, in the case of an infinitesimal time lapse ($\delta t<<1$) \ the
unitary operator $U(t)=e^{i\delta t\{c\mathbf{\alpha .p+}\beta
m_{0}c^{2}\}/\hbar }$\ can be approximated as:%
\begin{equation}
U(\eta )\simeq e^{i\delta t\{c\mathbf{\alpha .p}\}/\hbar }e^{i\delta t\beta
m_{0}c^{2}/\hbar }  \tag{A.14}
\end{equation}%
In configuration space this yields a displacement $\delta \mathbf{r=}<%
\mathbf{r}+\delta tc\mathbf{\alpha >=<r>}+(\delta t)\mathbf{v}_{gp}$ and a
phase shift $\Delta \varphi =\delta t<\beta >m_{0}c^{2}/\hbar $. For $\delta
t=\gamma \tau _{0}=\gamma h/m_{0}c^{2}$ (the boosted de Broglie period), the
phase shift is%
\begin{equation}
\Delta \varphi (\gamma h/m_{0}c^{2})=(\gamma
h/m_{0}c^{2})(m_{0}c^{2}/<H>)m_{0}c^{2}/\hbar =h/\hbar =2\pi  \tag{A.15}
\end{equation}%
These results are in agreement with the fact that the Hamiltonian is
actually the generator of the time development of a system described by a
wave packet. The approximate treatment provides only the displacement,
neglecting the dispersion of the wave packet.

\end{document}